\documentclass[]{aa}
\input epsfig.sty

\def\edoc{\end{document}}

\def\praz{\partial_{\eta}}
\def\pdwa{\partial_{\eta}^2}
\def\L3{\triangle}
\def\uu{u}
\def\X{\delta}
\def\Y{\Theta}
\def\P{p}
\def\T{\tau}
\def\Twave{{T}}
\def\stf{{time factor} }
\def\transfer{{\it  spectrum transfer function }}
\def\Dirac{\mbox{$\delta\hskip-.4em\delta$}}
\def\Lichnerowicz{{Darmois-Israel }}
\def\oraz{\& }
\def\tsigma{\Sigma}

\def\xx{{\bf x}}

\def\kk{{\bf k}}
\def\hh{{\bf h}}

\def\ue{{\uu}_{\kk(\epsilon)}}
\def\uerad{{\uu}_{\kk(\epsilon,1)}}
\def\uedust{{\uu}_{\kk(\epsilon,2)}}
\def\uv{{\uu}_{\kk(\vartheta)}}
\def\uvrad{{\uu}_{\kk(\vartheta,1)}}
\def\uvdust{{\uu}_{\kk(\vartheta,2)}}
\def\grant{No 2 P03D 014 17}

\renewcommand{\thesection}{\arabic{section}}
\renewcommand{\thesubsection}{\arabic{section} \arabic{subsection}}
\renewcommand{\theequation}{\arabic{equation}}
\begin{document}

\title{Acoustic instabilities at the transition from the radiation-dominated
       to the matter-dominated universe}

\author{Gra\.zyna Siemieniec-Ozi\c{e}b{\l}o
\and Andrzej Woszczyna}

\institute{Astronomical Observatory, Jagellonian University\\
Faculty of Mathematics, Physics and Computer Science\\
ul. Orla 171, 30--244 Krak\'ow, Poland}
\titlerunning{Acoustic instabilities in the universe}
\authorrunning{G. Siemieniec-Ozi\c{e}b{\l}o and A. Woszczyna}

\abstract{
The transition from acoustic noise in the radiation-dominated universe to the
density structures in the matter dominated epoch is considered. The initial
state is a stochastic field of sound waves moving in different directions. The
construction of the initial state is compatible with the hyperbolic type of
propagation equation for density perturbations, and parallel to the theory of
stochastic background of gravitational waves. Instantaneous transition between the
cosmological epochs is assumed, and \Lichnerowicz joining 
conditions
are applied to match solutions for sound waves with growing or decaying modes at the
decoupling. As a result a substantial amplification of the low scale structures
is obtained.}

\maketitle


\section{Introduction} 
Simple models of transitions between different cosmological epochs 
help us to understand amplification of scalar fields,
electromagnetic and gravitational waves in the expanding universe
(Frieman \& Turner 1984, \cite{Abbott&Harari,Hu&&,Grishchuk&&,Allen&Flanagan&Papa}). 
Typical
cosmology of that class consist of three phases: 1)~a semiclassical phase, which
is commonly identified with de Sitter stage, 2)~the radiation-dominated epoch
described by the equation of state $\P= \epsilon/3$, and 3)~the matter dominated
era when pressure is negligible $\P=0$. Transitions between them are assumed to be
instantaneous. The physical meaning of these models is close to that of simple
models of the particle scattering on rectangular potential barriers in quantum
mechanics. Asymptotic results weakly depend on the barrier profile, so we hope,
that in more realistic cosmology, the perturbation amplitude in the remote past
and the far future should only marginally depend on the transition details.

For scalar or electrodynamic fields, as well as for gravitational waves the
amplification can be measured by Bogolubov coefficients (\cite{Birrell&Davies}).
These fields are governed by hyperbolic partial differential equations in each
of discussed epochs and the changes in the equation of state - the change in the
background dynamics - result in a specific relations between Fourier modes in
the {\it in} and {\it out} state, respectively. For some
frequencies, counter-propagating waves are substantially amplified (creation of
pairs of particles with opposite momentum) - the so called back scattering
effect (\cite{Parker&&}) comes into play\footnote{For low frequency gravitational waves
standing waves of substantial amplitude may appear (squeezed state of
gravitational field (\cite{Grishchuk&&}))}. Similar effects may occur in acoustic
field (\cite{Lukash99&&}). Fields enhancement on non static background is generally
referred to parametric amplification (\cite{Grishchuk2&&}).

Neither parametric amplification nor particle creation theories provide
appropriate language to describe the growth of density perturbations
in the transition to matter-dominated epoch. The scalar perturbations form waves
(\cite{Sachs&Wolfe}, \cite{White_PC&&}, \cite{Field&Shepley}) 
(in quantum phonon--approach \cite{Lukash&&}, \cite{Chibisov&Mukhanov}) 
only in the epochs of non vanishing pressure, but they
transform into non-travelling ingomogeneities (growing and decaying modes) in
the matter dominated epoch ($\P=0$).  The change then is two-fold: 1) the
transition modifies the background dynamics, 2) the propagation equation change
its differential type. Bogolubov coefficients loose
their physical meaning. Yet, the general scheme of the field propagation
throughout the transition epoch is the same: for the scalar or electrodynamic
fields the continuity of each field and its time derivative must be satisfied at
the transition, for metric perturbations (both scalar and tensor) the first and
the second fundamental forms must be continuous (\Lichnerowicz conditions).

Classical perturbations in the cosmological models with sharp transitions  have
been investigated by Kodama and Sasaki (\cite{Kodama&Sasaki}) or Hwang and
Vishniac (\cite{Hwang&Vishniac}). These authors limit themselves to the standing
wave solutions and to regime of low frequencies. A similar task for a universe
with the radiation and dust mixture has been undertaken by Mukhanov, Feldman and
Brandenberger (\cite{Mukhanov&Feldman&Brandenberger}) also in the long wave limit.
In quantum theories the radiation to matter transition has been extensively
discussed by Grishchuk (\cite{Grishchuk1&&}) as a part of a more complex
cosmological model including an inflationary stage. Large scale perturbations
where investigated there in the context CMBR temperature fluctuations. Relatively little is known
about the low scale inhomogeneities.

In this paper we present exact formulae for density perturbations in the
universe with sharp transition in a full range of frequencies. We start with
acoustic noise in the radiation era and investigate its
transition to density structures in the matter dominated epoch. We employ the
autocorrelation function as a measure of structure, or equivalently its Fourier
transform--the spatial spectrum of inhomogeneities. This measure  agrees well
with what other authors propose in cosmology. However, despite the cosmological
practice we do not limit the basis of elementary solutions to "growing modes".
Instead, we take into account the complete basis of Fourier modes, adequate to
the hyperbolic character of the propagation equation and in full analogy to a
stochastic description of gravitational waves (\cite{Allen&Flanagan&Papa}) and
quantum (phonon) theories (\cite{Lukash&&}, \cite{Chibisov&Mukhanov}, 
\cite{Grishchuk1&&}). Since we disregard inflation, there is no squeezing mechanism
(\cite{Grishchuk2&&}) for the acoustic field. We examine the spatial power
spectrum without being limited to standing waves, and show that substantial
amplification occurs in the high frequency regime, which has not been
investigated till now.

The structure of the paper is the following: in Section~\ref{RW} we discuss
exact solutions to the perturbation equations in both radiation and matter eras
employing orthogonal gauge. In the Section~\ref{szycie} we construct the 
2-epoch model with the instantaneous transition in the equation of state, by use
of the \Lichnerowicz conditions. The Section~\ref{random} is devoted to the
stochastic properties of the acoustic field. Finally, we investigate the time
evolution of the spatial spectrum of cosmological inhomogeneities.

\section{Perturbations in the Robertson-Walker universe}
\label{RW}

There are several independent methods to describe scalar perturbations in a
gauge-invariant way 
(\cite{Olson&&},        
\cite{Bardeen&&},    
\cite{Brandenberger&Kahn&Press}, 
\cite{Lyth&Stewart}, 
\cite{Ellis&Bruni}).
For the universe filled
with an ideal fluid\footnote{With the diagonal energy momentum tensor
$T_{\mu\nu}=(\epsilon+\P)u^{\mu}u^{\nu}+\P g_{\mu\nu}$} with arbitrary non-
vanishing pressure $\P$ all of them lead to the same result: under appropriate choice
of the perturbation variables, the propagation equations converge to the wave
equation (\cite{Sachs&Wolfe}, \cite{Field&Shepley}, \cite{Chibisov&Mukhanov}, \cite{Golda&Woszczyna2}). 
The density perturbations form travelling waves. Below we
consider rotationless fluid, hence the hypersurfaces orthogonal to the fluid
flow can be globally defined, and therefore, the orthogonal gauge 
(\cite{Lyth&Mukherjee}, \cite{Lyth&Stewart}, \cite{Padmanabhan&&}) 
is naturally applied.

The linear corrections to the energy density and the expansion rate
evolve according to
	\begin{eqnarray}
	\label{eq1}
 \partial_{t}\delta\epsilon(t,\xx)&=&
 -\P_0\delta\vartheta(t,\xx)
 - \epsilon_0\delta\vartheta(t,\xx)
 - \vartheta_0\delta \epsilon(t,\xx)\\
 	\label{eq2}
 \partial_{t}\delta\vartheta(t,\xx)&=&
 -4 G \pi\delta\epsilon(t,\xx)
 -\frac{\nabla^2\delta \P(t,\xx)}{\P_0+\epsilon_0}\\
 &-&\frac{2}{3} \vartheta_0\delta \vartheta (t,\xx)
 -\frac{ (24 G \pi\epsilon_0-{\vartheta_0^2})\delta \P(t,\xx)}{3(\P_0+\epsilon_0)}
 	\nonumber\end{eqnarray}
where the subscript ``0" refers to the unperturbed, background FRW universe 
($\epsilon_0= \epsilon_0(t)$, and $\P_0=\P_0(t)$ are solely functions of time). The
equations (\ref{eq1}), (\ref{eq2}) are derived from the Raychaudhuri and the
continuity equations by use of the standard linearization procedure and by
replacing the proper time by the orthogonal time. {Below we limit 
ourselves to spatially flat universe $K=0$. In this case the background energy density 
and the expansion rate are related to each other by $8 \pi G\epsilon_0=\vartheta_0^2/3$.}

Let us assume now that the universe is filled by the single fluid with the
equation of state $P/\epsilon =w=\mbox{constant}$, where the value of $w$
determines both the evolution of the background metric (unperturbed universe) and
the sound velocity for small perturbations. We express the energy density and the
expansion rate as the composition of the background energy values $\epsilon_0$,
$\vartheta_0$ and the small inhomogeneous correction $\delta\epsilon= \epsilon_0
\X(t,\xx)$, $\delta \vartheta={\vartheta }_0 \Y(t,\xx)$
	\begin{eqnarray}
	\label{eq3.1}
\epsilon (t,\xx)&=&\epsilon_0(t)(1+\X(t,\xx))\\
	\label{eq3.2}
\vartheta (t,\xx)&=&\vartheta_0(t)(1+\Y(t,\xx))
	\end{eqnarray}
where $\X(t,\xx)$ and $\Y(t,\xx)$ play the role of the density and expansion contrasts, respectively.
Transforming system (\ref{eq1}), (\ref{eq2}) to a second order propagation
equation for~$\X$ we obtain a partial differential equation of the form
	\begin{eqnarray}
w \L3 \X(t,\xx)
&=&\frac{\vartheta_0^2}{6} (w-1) (1+3w) \X(t,\xx)\nonumber\\
&+&\vartheta_0\left(\frac{2}{3}-w\right) \partial_{t}\X(t,\xx)
+\partial_{t}^2 \X(t,\xx)
	\label{eq4}\end{eqnarray}
with the background evolution given by $\vartheta_0(t)=\frac{2}{(1+w)t}$ or
equivalently in the conformal time $\eta$
	\begin{eqnarray}
w\L3 \X(\eta,\xx)
&=&\left(\frac{3w-3}{3w+1}\right)\frac{2}{\eta^2}\X(\eta,\xx)\nonumber\\
&-&\left(\frac{3w-1}{3w+1}\right)\frac{2}{\eta}\praz \X(\eta,\xx)
+\pdwa \X(\eta,\xx)
	\label{eq5}\end{eqnarray}
The conformal time $\eta$ is defined here as the integral $\int \frac{1}{a(t)}dt$
of the scale factor reciprocal over the orthogonal time $t$ and 
$\L3$
stands for the Laplace operator in 3-dimensional Euclidean space.
Equation (\ref{eq5}) can be solved analytically by use of Fourier transform.
Actually, we are interested in two special cases $w= 1/3$ (radiation- filled
universe) and $w=0$ (matter-domination epoch).

The propagation equations for small perturbations (\ref{eq4}), (\ref{eq5}) (and
consequently (\ref{eq6.1}) below) do not contain the gravitational constant
$G$, which means that inhomogeneities do not self-gravitate unless the linear
regime breaks down. They evolve as acoustic waves in the expanding gas medium
(compare \cite{Sachs&Wolfe}, \cite{Stone&&})). All the perturbation equations
obtained in different gauge invariant formalisms can be reduced to an equation of
the form (\ref{eq4}) by suitable changes of variables (and with different
meaning of the variable $\X$ and the time parameter) (\cite{Golda&Woszczyna2}).
The necessary changes reads (in the notation: $\mbox{\it reference:
original notation\/}\to \X$)
\noindent
(\cite{Sakai&&}): $K\to \X$;\\
(\cite{Bardeen&&}): $\rho_{\rm m}\to \X$;\\
(\cite{Kodama&Sasaki} chap. IV):  ${\mit\Delta}\to \X$;\\
(\cite{Lyth&Mukherjee}): $\delta\to \X$;\\
(\cite{Padmanabhan&&}): $\delta\to \X$;\\
(\cite{Brandenberger&Kahn&Press}): ${\mit\Phi}_H\slash\rho {a^2} \to \X$;\\
(\cite{Ellis&Bruni&Hwang}): ${\cal D}\to \X$.\\
\noindent
Transformations of these equations to conformal time (if parameterized
differently) are necessary.

In the universe filled with highly relativistic matter, the scale factor $a(\eta)$
evolves as a linear function of the conformal time: $a(\eta)=\sqrt{{\cal M} /3}
\,\eta$, and preserve ${\cal M} =\epsilon_0 a^4$ as the constant of motion.  
Equation (\ref{eq5}) expressed in conformal time takes the canonical form
(independent of first derivatives)
	\begin{equation}
	\label{eq6.1}
\frac{1}{3}\L3 \X(\eta,\xx)
=-\frac{2\X(\eta,\xx)}{{\eta }^2}
+ \pdwa \X(\eta,\xx)
	\end{equation}
Equation (\ref{eq6.1})  is essentially the same as the propagation
equation for gravitational waves in the dust-filled universe 
(\cite{Grishchuk0&&}, \cite{White&&}). It reduces to the wave equation in its normal form
	\begin{equation}
	\label{eq6.2}
\frac{1}{3}\L3 \widehat{\X}(\eta ,\xx)
=\pdwa\widehat{\X}(\eta ,\xx)
	\end{equation}
for $\widehat{\X}(\eta ,\xx)$ defined as
	\begin{equation}
	\label{eq7.1}
\widehat{\X}(\eta ,\xx)
=\frac{1}{\eta } \praz (\eta \X(\eta, \xx)).
	\end{equation}
The variable $\widehat{\X}(\eta ,\xx)$ is the orthogonal-gauge analogue to the Field-
Shepley variable $H$  (\cite{Field&Shepley}), or the Sachs-Wolfe variable $E$ 
\cite{Sachs&Wolfe}. Solutions $\widehat{\X}
(\eta,\xx)$ and $\X(\eta,\xx)$ expand into Fourier series in the way appropriate for
massless scalar fields (\cite{Birrell&Davies}),
	\begin{eqnarray}
	\label{eq9.1}
\widehat{\X}(\eta,\xx)
&=&\int ({\sf A}_{\kk} {\sf u}_{\kk(\epsilon)}(\eta,\xx)
+{\sf A}_{\kk}^*{\sf u}_{\kk(\epsilon)}^*(\eta,\xx))d\kk\\
	\label{eq9.2}
\X(\eta,\xx)
&=&\int ({\cal A}_{\kk} \ue (\eta,\xx)
+{\cal A}_{\kk}^* \ue ^* (\eta,\xx))d\kk
	\end{eqnarray}
The modes ${\sf u}_{\kk(\epsilon)}(\eta ,\xx)$ are simply $\frac{1}{\sqrt{2\omega}}
e^{i(\kk \xx-\omega  \eta )}$, while $\ue (\eta,\xx)$ can be found as
	\begin{eqnarray}
\ue (\eta,\xx)
&=&\frac{1}{\eta}\int \eta{\sf u}_{\kk(\epsilon)}(\eta ,\xx)d\eta\nonumber\\
&=&\frac{1}{\sqrt{2 \omega}}\left(1+\frac{1}{i\omega\eta}\right){e^{i(\kk \xx-\omega \eta)}}
	\label{eq10}\end{eqnarray}
The generic perturbation $\X(\eta,\xx)$ is composed of travelling plane waves $\ue $
with decreasing amplitude. (Similar solutions are known in the theory of
gravitational waves (\cite{White&&}), and scalar field
\cite{Stebbins&Veerarghavan}.). The  Fourier coefficient ${\sf A}_{\kk}=-i\omega {\cal A }
_\kk$ is an arbitrary complex function of the wave number $\kk$, while $\X(\eta,\xx)$
keeps real values. The frequency $\omega$  obeys the dispersion relation
$\omega^2= k^2/3$, hence waves of all length-scales propagate with the same phase and 
group velocity\footnote{This is not the case of open or closed universes,
where sound are dispersed on the space curvature (\cite{Golda&Woszczyna2}).}
(compare \cite{Chibisov&Mukhanov, Sachs&Wolfe, White_PC&&}). Modes
$\ue $, although different from the simple eikonal form, are still
orthonormal in the sense of the Klein-Gordon scalar product.

In the epoch of matter dominance $m =\epsilon_0 a^3$ is the constant of
motion and the scale factor evolves as $a(\eta)=\frac{m\eta^2}{12}$. The
propagation equation expressed in orthogonal gauge (in all formalism mentioned
above) reads
	\begin{equation}
	\label{eq11}
-\frac{6}{\eta^2}\X(\eta,\xx)
+\frac{2}{\eta}\praz \X(\eta,\xx)
+\pdwa \X(\eta,\xx)=0.
	\end{equation}
Vanishing pressure implies the absence of the Laplace operator, consequently,
the general solution consists of growing and decaying solutions involving two 
arbitrary functions of the space coordinates $f_1(\xx)$ and $f_2(\xx)$:
	\begin{equation}
	\label{eq12}
\X(\eta,\xx)
={f_1}(\xx) \eta^2
+f_2(\xx)\eta^{-3}
	\end{equation}
This solution expands into Fourier series
	\begin{eqnarray}
\X(\eta,\xx)
&=&\int(a_\kk e^{i \kk \xx}+a_\kk^* e^{-i \kk \xx})\eta^2
+(b_\kk e^{i \kk \xx} +b_\kk^* e^{-i \kk \xx})\eta^{-3} d\kk\nonumber\\
&&
	\label{eq13}\end{eqnarray}
where the coefficients $a_\kk$ and $b_\kk$ are arbitrary complex functions of $\kk$.

\section{Matching conditions in the transition epoch}
\label{szycie}

Consider now the two-epoch cosmological model composed of both, the radiation epoch
(governed by the equation of state $\P=\epsilon /3$), and the succeeding epoch of 
matter domination (with $\P=0$). Below, the quantities related to these two epochs
will appear with the indices (1) and (2), respectively. We assume that the
transition between the epochs is instantaneous and occurs on the hypersurface $\Sigma$
orthogonal to the four velocity $u^\mu$ of the matter
content.

The initial Cauchy conditions are unique and consistent on the hypersurface
$\Sigma$ if the first and the second fundamental forms are equal
(\cite{Darmois&&, Hawking&Ellis})
	\begin{eqnarray}
	\label{eq14}
h_{\mu\nu\,(1)}(\Sigma)&=&h_{\mu\nu\,(2)}(\Sigma),\\
	\label{eq15}
\chi_{\mu\nu\,(1)}(\Sigma)&=&\chi_{\mu\nu\,(2)}(\Sigma).
	\end{eqnarray}
Subscripts (1) and (2) refer to both half-spaces divides by the surface $\Sigma$. For
unperturbed background these equations imply the continuity of  the scale factor $a(\eta)$
and its first derivative
	\begin{eqnarray}
	\label{eq16}
a_{(1)}(\Sigma)&=&a_{(2)}(\Sigma),\\
	\label{eq17}
\partial_\eta a_{(1)}(\Sigma)&=&\partial_\eta a_{(2)}(\Sigma).
	\end{eqnarray}
In our two-epoch model the matching conditions are satisfied by
	\begin{eqnarray}
	\label{eq18}
a_{(1)}(\eta)&=&\sqrt{\frac{\cal M}{3}}\,\eta,\\
	\label{eq19}
a_{(2)}(\eta)&=&\frac{\sqrt{\cal M}(\eta
+\eta_{\tsigma})^2}{4\sqrt3\eta_{\tsigma}},
	\end{eqnarray}
where $\eta_{\tsigma}$ denotes the time of the transition, and $\cal M$ is the
constant of motion ${\cal M} =\epsilon_0 a_{(1)}^4$ of the radiation filled
universe.

In the gauge-orthogonal formalism conditions (\ref{eq14}, \ref{eq15})
can be rewritten to directly join the density and expansion perturbations in both
epochs, before and after the transition. From (\ref{eq14}, \ref{eq15}) one can
easily find that the energy density $\epsilon$ and the expansion rate
$\vartheta$ are continuous on $\Sigma$. Indeed, the energy density $\epsilon$ is
related to the induced curvature $R^{(3)}$ and the second fundamental form by
(\cite{Hawking&Ellis})
	\begin{equation}
	\label{eq20}
2\epsilon
=R^{(3)}-(\chi^\mu_\mu)^2-\chi_{\mu\nu} \chi^{\mu\nu}
	\end{equation}
The Ricci scalar $R^{(3)}$ on $\Sigma$ consists of the metric form $h_{\mu \nu}
(\Sigma)$ and its {\it space\/} derivatives, so $R^{(3)}$ like $h_{\mu \nu}
(\Sigma)$, is continuous in the transition. $\chi_{\mu\nu}$ is continuous on the
strength of (\ref{eq15}). As a consequence, (\ref{eq20}) ensures, that $\epsilon$
is continuous. On the other hand the continuity of the expansion rate
$\vartheta(t)$ comes directly from the same property of the second fundamental
form $\chi_{\mu\nu}$. Eventually, after {eliminating} unperturbed values $\epsilon_0$
and $\vartheta_0$, the density perturbations obey
	\begin{eqnarray}
	\label{eq21}
\delta\epsilon_{(1)}(\Sigma)&=&\delta\epsilon_{(2)}(\Sigma),\\
	\label{eq22}
\delta\vartheta_{(1)}(\Sigma)&=&\delta\vartheta_{(2)}(\Sigma)
	\end{eqnarray}
where
	\begin{eqnarray}
	\label{eq23}
\delta\epsilon(\eta,\xx)&=&\epsilon_0 \X(\eta,\xx),\\
	\label{eq24}
\delta\vartheta(\eta,\xx)&=&-\frac{\vartheta_0}{{P_0}+\epsilon_0}
\delta\epsilon(\eta ,\xx)-
\frac{1}{a (P_0+\epsilon_0)} \partial_\eta
\delta\epsilon(\eta,\xx),
	\end{eqnarray}
as derived from (\ref{eq1}) and (\ref{eq3.1}). It is important to note that the
time derivatives of $\epsilon$ and $\vartheta$ may jump, therefore neither $\delta
\epsilon'(\eta)$ nor $\X'(\eta)=(\delta \epsilon(\eta)/\epsilon)'$ are continuous
on $\Sigma$.

The starting point for the investigations of acoustic fields and the structures they produce
in the two-epoch universe is the analysis of a single monochromatic wave (a
single Fourier mode)
	\begin{equation}
	\label{eq25}
\uerad =\frac{1}{\sqrt{2\omega}}\left(1+\frac{1}{i\omega
\eta}\right) e^{i(\kk \xx-\omega\eta)}.
	\end{equation}
This wave, while falling on $\Sigma$, generates on the ``other side'' of the
transition surface a mixture of growing and decaying mode
	\begin{equation}
	\label{eq26}
\uedust =a_\kk\left(\frac{\eta
+\eta_{\tsigma}}{\eta_{\tsigma}}\right)^2 e^{i \kk\xx}
+b_\kk\left(\frac{\eta +\eta_{\tsigma}}{\eta_{\tsigma}}\right)^{-3}e^{i \kk \xx}.
	\end{equation}
The two coefficients $a_\kk$ and $b_\kk$ are uniquely determined from (\ref{eq21},
\ref{eq22}). Indeed, evaluating (\ref{eq23}, \ref{eq24}) for the perturbation
(\ref{eq25}, \ref{eq26}) and imposing the matching condition (\ref{eq21},
\ref{eq22}) one obtains
	\begin{eqnarray}
	\label{eq27}
a_\kk	&=&-\frac{3}{40}\frac{1}{\sqrt{2\omega}}
i\omega\eta_{\tsigma} e^{-i\omega\eta_{\tsigma}}\\
	\label{eq28}
b_\kk	&=&\frac{1}{\sqrt{2\omega}}\left(8+\frac{8}{i\omega\eta_{\tsigma}}+
\frac{12}{5}i\omega\eta_{\tsigma}\right)e^{-i\omega\eta_{\tsigma}}
	\end{eqnarray}
The frequency $\omega$ refers to the acoustic wave in the radiation epoch, and
is related to the wave number $k$ by the linear dispersion relation $\omega=
k/\sqrt3$. In the epoch of matter dominance, the perturbations lose their wave
character, therefore it may be better to parameterise them by the wave number
$k$, which has a well defined meaning in both epochs. Now, the modes $\ue$
with coefficients (\ref{eq27}, {\ref{eq28}) take the form
	\begin{eqnarray}
	\label{eq29.1}
\uerad 
&=&\frac{3^{1/4}}{\sqrt{2k}}
\left(1+\frac{\sqrt3}{ik\eta}\right)e^{i\kk \xx-\frac{ik\eta}{\sqrt{3}}}\\
	\nonumber
\uedust 
&=&\frac{3^{1/4}}{\sqrt{2k}}e^{i\kk \xx-\frac{ik\eta_{\tsigma}}{\sqrt{3}}}
\left(-\frac{3}{40}\frac{i k\eta_{\tsigma}}{\sqrt{3}}\left(1+\frac{\eta}{\eta_{\tsigma}}\right)^2\right.\\
&+&\left.\left(8+\frac{8\sqrt{3}}{ik\eta_{\tsigma}}+\frac{12}{5}
\frac{i k\eta_{\tsigma}}{\sqrt{3}}\right)\left(1+\frac{\eta}{\eta_{\tsigma}}\right)^{-3}
\right) 
	\label{eq29.2}
	\end{eqnarray}
The second independent solution consists of complex conjugates of
$\uerad $ and $\uedust $. We will include it later to
restore the complete Fourier basis.

\section{Random acoustic fields}
\label{random}

The acoustic field in the early universe is shaped by thermodynamic or quantum
phenomena acting prior to and during the radiation era. Their probabilistic
nature leads to stochastic description. We restrict ourselves to stochastic
processes homogeneous in the {\it broad sense\/}, called also {\it weakly
homogeneous processes\/}, which  keep their mean value and variance (standard
deviation) invariant under translations. The two-point autocorrelation functions
for them are functions of the distance between points solely --- not of these
points' positions  (\cite{Loeve&&}).
Processes homogeneous in the {\it broad sense\/} have their Fourier
representations (\cite{Loeve&&, Sobczyk&&, Yaglom&&}) 
	\begin{equation}
\widehat{\X}(\eta,\xx)
=\int ({\sf A}_{\kk} {\sf u}_{\kk(\epsilon)}(\eta,\xx)
+{\sf A}_{\kk}^*{\sf u}_{\kk(\epsilon)}^*(\eta,\xx))d\kk
	\end{equation}
where the integral is understood to be the stochastic integral, and the Fourier
coefficients ${\sf A}_{\kk}$ are random variables. Their expectation values
fulfil  (\cite{Sobczyk&&, Yaglom&&})
	\begin{eqnarray}
	\label{eq30.1}
E[{\sf A}_{\kk}{\sf A}_{k'}^*]	&\sim&\Dirac(\kk-\kk')\\
	\label{eq30.2}
E[{\sf A}_{\kk}{\sf A}_{k'}]		&=&0
	\end{eqnarray}
where $\Dirac$ denotes Dirac's delta. Conversely, each process obeying (\ref{eq30.1}, \ref{eq30.2}) is homogeneous in
the broad sense.

The relations (\ref{eq30.1}, \ref{eq30.2}) have a clear physical meaning. The first
of them expresses the statistical independence of waves with different wave-vectors.
The second means that phases of perturbations at any moment and any place are
statistically independent. Altogether they assure statistical independence of
waves moving in different directions.
This stochastic process can be also expressed in terms of the $\X(\eta,\xx)$ variable. With
help of the random Fourier coefficients ${\cal A}_{\kk}$ satisfying
	\begin{eqnarray}
	\label{eq31.1}
E[{\cal A}_{\kk}{\cal A}_{\kk'}^*]	&=&P_k\,\Dirac(\kk-\kk')\\
	\label{eq31.2}
E[{\cal A}_{\kk}{\cal A}_{\kk'}^*]	&=&0
	\end{eqnarray}
we write
	\begin{equation}
	\label{eq32}
\X[\eta ,\xx]=\int \left({\cal A}_{\kk}\ue (\eta,\xx) +
{\cal A}_{\kk}^*\ue ^*(\eta,\xx)\right)d\kk
	\end{equation}
We assume that the power spectrum of the acoustic noise $P_k$ depends solely on
the magnitude of the wave vector and not on its direction.  The perturbation is
generic --- no additional constraints, nor any ordering or squeezing mechanisms have been
introduced. The construction of a random acoustic field is identical with the
construction of random field of gravitational waves in the matter dominated
universe (\cite{Abbott&Harari, Allen&&, Allen&Romano, Maggiore&&}).
It reflects
the equivalence of the propagation equations for both these classes of
perturbation. Stochastic acoustic field fulfilling conditions (\ref{eq31.1}-\ref{eq31.2})
can be understood as a classical limit of the quantum phonon approach 
(\cite{Lukash&&, Chibisov&Mukhanov, Grishchuk1&&}).

Modes $\ue $ while regularly extended across the transition
hypersurface $\Sigma$ define a stochastic structure (by means of integral
(\ref{eq32})) in the matter dominated universe. The commonly used measure of this
structure is the two-point autocorrelation function
	\begin{equation}
	\label{eqRdens}
R(\eta,h)=\frac{1}{4\pi}\int E[{\X}(\eta,\xx){\X}
(\eta,\xx+\hh)] \Dirac(\hh{\cdot}\hh-1)\,{\rm d}\hh
         \end{equation}
defined on the constant time hypersurfaces. Under conditions
(\ref{eq30.1}--\ref{eq31.2}) this function reads
\begin{eqnarray}
R(\eta,h)&=&     \frac{1}{4\pi}\int 2u_{\kk}u_{\kk}^*{\cal P}_k
                \exp(i\kk{\cdot}\hh)\Dirac(\hh{\cdot}\hh-1)\,
                {\rm d}\kk\,{\rm d}\hh\nonumber\\
       	&=&	\int_0^\infty 4\pi k^2
\frac{\sin(hk)}{hk} p_k(\eta)\,{\rm d} k
\end{eqnarray}
where
	\begin{equation}
	\label{eq35}
p_k(\eta)=2 \ue \ue ^* P_k=\T_{(\epsilon)} P_k
	\end{equation}
plays the role of the structure spatial spectrum and $\ue (\eta,\xx)$ are
modes defined by (\ref{eq29.1}, \ref{eq29.2}) extended to both epochs before and
after the transition. In this way acoustic noise in the radiation era with a given
spectrum $P_k$ determines uniquely (via \Lichnerowicz conditions) the spatial
spectrum of inhomogeneities $p_k(\eta)$ at any stage of the structure formation
process.

{It is important to clearly distinguish between $p_k$ and $P_k$, and understand
their roles in the cosmological context. As defined above, $p_k$ is the Fourier transform of the
two-point autocorrelation function.
In the cosmological literature it is called the {\em power spectrum}\footnote{In this
context we consequently use the name {\em spatial power spectrum}.} by analogy to
similar concept known in the analysis of time series (\cite{Anderson&&}), 
but its physical dimension is different.
On the other hand, $P_k$ defined by (\ref{eq31.1}) is the genuine power spectrum of
the acoustic field (with the same physical sense as the Planck power spectrum 
of the electromagnetic radiation),
and can be obtained from Hamiltonian description (\cite{Lukash&&}). Although in
cosmology one cannot directly observe $P_k$, this quantity defines the physical
state of acoustic field. This is $P_k$ not $p_k$, which should be either guessed, or
inferred from fundamental laws of physics (\cite{Chibisov&Mukhanov}). The shape of
$P_k$ is not precised in this paper.}

The \stf $\T_{(\epsilon)}=2 \ue \ue ^*$ converts the acoustic spectrum $P_k$
into the spatial-spectrum $p_k(\eta)$. (It is numerically equal to $p_k(\eta)$ 
for the white noise acoustic field: $P_k=const$.) Factor
$\T_{(\epsilon)}$ contains the entire time dependence of cosmological
inhomogeneities. Employing extended modes (\ref{eq29.1}, \ref{eq29.2}), we find
	\begin{eqnarray}
\T_{(\epsilon,1)}
&=&\frac{\sqrt{3}}{k}\left(1+\frac{1}{\tilde{k^2}\tilde{\eta}^2}\right)
\label{eq37.1}\\
	\nonumber
\T_{(\epsilon,2)}
&=&\frac{\sqrt{3}}{k}\left(\frac{128}{5}\left(1+\frac{5}{2\tilde{k}^2}
+\frac{9\tilde{k}^2}{40}\right)(1+\tilde{\eta})^{-6}\right.\nonumber\\
&+&\frac{3(10-3\tilde{k}^2)}{25}(1+\tilde{\eta})^{-1}
+\left.\left(\frac{3\tilde{k}}{40}\right)^2(1+\tilde{\eta})^4\right).
\label{eq37.2}
	\end{eqnarray}
where $\tilde{\eta}$ is the normalized time parameter
$\tilde{\eta}= \eta/\eta_{\tsigma}$ and the $\tilde{k}$ is the modified wave number
$\tilde{k}= k\eta_{\tsigma}/\sqrt{3}$ which measures the number of oscillations
within the sound horizon on $\Sigma$. We hold
the factor $\sqrt{3}/k$ in both $\T_{(\epsilon,1)}$ and $\T_{(\epsilon,2)}$ to
keep modes $\uerad $ orthonormal in the sense of the Klein-Gordon scalar product
in the radiation era. It is easy to check that the \stf $\T$ is
continuous on $\Sigma$.

In the radiation era, perturbations greater than the sound horizon decrease,
while the low scale ones maintain constant amplitude. This can be seen directly
from (\ref{eq37.1}). The result qualitatively agrees with quantum theories (see
\cite{Mukhanov&Feldman&Brandenberger} part III, formula 20.6). Quantitative
difference comes from the different gauge choice\footnote{The quadratic behaviour
of the decreasing term is characteristic for the orthogonal gauge. In the
synchronous system of reference this term decays as
$\frac{1}{\tilde{k}^4\tilde{\eta}^4}$. It can be easily derived in the Field--Shepley 
formalism, see formula (5.3) of (\cite{Chibisov&Mukhanov}) after
substituting solutions of~(4.7) and evaluating the integral over $\eta$.
Equivalently, it can be proved directly in the original Lifshitz formalism 
(\cite{Golda&Woszczyna1}).}. It should be emphasized that the spectrum $P_k$, the
spatial spectrum $p_k$ and the parameter $\T$ are invariant under unitary
transformations. Therefore, the result is physically well defined -- it is not the
effect of any particular choice of the Fourier basis.

In the matter era the perturbations (\ref{eq29.2}) are a specific mixture
of growing and decaying solution (compare 
\cite{Ellis&Hellaby&Matravers, Liang&&}). Therefore, despite vanishing 
pressure, their evolution
depend on their length-scales. The dominant growing term 
$(1+\tilde{\eta})^4$ in (\ref{eq37.2}) is multiplied by $\tilde{k}^2$, which
means strong amplification of short waves.
}

A peculiar velocity field can be measured by the deviation $\delta\vartheta$
from the homogeneous Hubble flow $\vartheta_0=3H$. Modes of the fluid
compression\footnote{The orthogonal gauge realize the comoving system of
reference, where the four velocity is globally chosen as $u=(1,0,0,0)$,
therefore one cannot describe the peculiar velocity field directly by $\delta u$}  are
associated to the density modes (\ref{eq10}), (\ref{eq25}). On the strength of 
(\ref{eq24}) we obtain
	\begin{equation}
	\label{eq38}
\delta\vartheta_{(1)}
=\frac{1}{\sqrt{2\omega}}\frac{3\sqrt3}{2\sqrt{\cal M}\eta^2}
\left(1+\frac{1}{i\eta\omega}+\frac{i\eta\omega}{2}\right)e^{i(\kk \xx-\eta\omega)}.
	\end{equation}
$\delta\vartheta$ play a role in the matching conditions, as the \Lichnerowicz
conditions require that both the density and expansion fields be continuous at
the transition. (Their time derivatives my undergo discontinuity.) The expansion
contrast $\Y= \delta\vartheta/\vartheta_0$ can be expressed as a stochastic integral
	\begin{equation}
	\label{eq39.0}
\Y(\eta ,\xx)
=\int({\cal A}_{\kk} \uv (\eta,\xx)
+{\cal A}_{\kk}^* \uv ^*(\eta,\xx))d\kk
	\end{equation}
with the same coefficients ${\cal A}_{\kk}$ as in (\ref{eq9.2}), but with a different
modes $\uv $, given by
	\begin{eqnarray}
	\label{eq39.1}
\uvrad 
&=&\frac{3^{1/4}}{2\sqrt{2k}}
\left(1+\frac{\sqrt{3}}{ik\eta}+\frac{1}{2}\frac{ik\eta}{\sqrt{3}}\right)
e^{i\kk \xx- ik\frac{\eta}{\sqrt{3}}},\\
	\label{eq39.2}
\uvdust
&=&\frac{3^{1/4}}{\sqrt{2 k}}e^{i\kk \xx-\frac{ik\eta_{\tsigma}}{\sqrt{3}}}
\left(\frac{ik\eta_{\tsigma}}{40\sqrt{3}}
\left(1+\frac{\eta}{\eta_{\tsigma}}\right)^2\right.\nonumber\\
&+&\left.\left(4+\frac{4\sqrt{3}}{ik\eta_{\tsigma}}
+\frac{6 ik\eta_{\tsigma}}{5\sqrt{3}}\right)
\left(1+\frac{\eta}{\eta_{\tsigma}}\right)^{-3}\right).
	\end{eqnarray}
In close analogy with the density perturbations, one can express the peculiar
velocities (perturbation in the expansion rate) in terms of the spatial
autocorrelation or spectrum. This time 
	\begin{equation}
	\label{eqRexp}
R(\eta,h)=\frac{1}{4\pi}\int E[{\Y}(\eta,\xx){\Y}
[\xx+\hh,\eta)] \Dirac(\hh{\cdot}\hh-1)\,{\rm d}\hh
         \end{equation}
and the \stf $\T$ is given by formula (\ref{eq35}) after replacing
$\ue $ by $\uv $
	\begin{eqnarray}
	\label{eq40.1}
\T_{(\vartheta,1)}&=&\frac{\sqrt{3}}{4k}
\left(\frac{\tilde{\eta}^2\tilde{k}^2}{4}
+\frac{1}{\tilde{\eta}^2\tilde{k}^2}\right),\\
	\nonumber
\T_{(\vartheta,2)}&=&\frac{\sqrt{3}}{k}
\left(\frac{32}{5}\left(1+\frac{5}{2\tilde{k}^2}
+\frac{9\tilde{k}^2}{40}\right) (1+\tilde{\eta})^{-6}\right.\nonumber\\
&-&\left.\frac{(10-3\tilde{k}^2)}{50} (1+\tilde{\eta})^{-1}
+\left(\frac{\tilde{k}}{40}\right)^2 (1+\tilde{\eta})^4\right).
	\label{eq40.2}
	\end{eqnarray}
Unlike the density spectrum, the spectrum of the expansion rate monotonically
increases in the radiation dominated universe (\ref{eq40.1}, see also the
behaviour of modes $\uvrad $ --- \ref{eq39.1}). This is so, because the decrease
in peculiar velocities is less than the decrease in homogeneous Hubble flow.
Increase of the expansion contrast $\Y$ with time does not invalidate the
acoustic approximation, as long as the gas velocities are small when compare
with the sound velocity (\cite{Whitham&&}). In our case the acoustic approximation
is valid when the contribution of terms quadratic in $\delta\vartheta(\eta)$ to
$\delta\vartheta'(t)$ (\ref{eq2}) is negligible. For large values of time this is
assured by (\ref{eq38}). Therefore, a perturbation initiated as an acoustic perturbation 
remains acoustic during the entire radiation era, even though
$\delta\vartheta/\vartheta$ increases.

\begin{figure}[h]
\centering 
\psfig{file=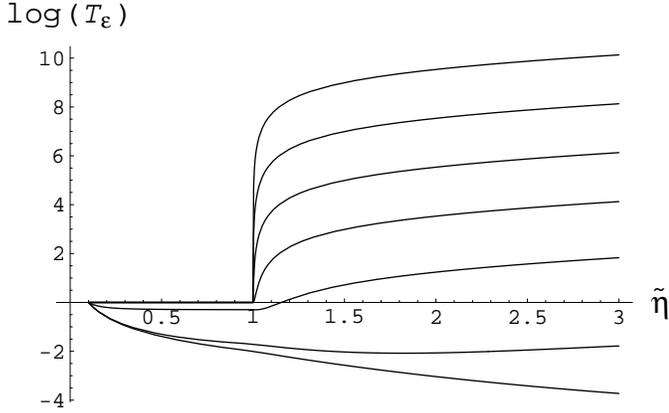,width=88mm 
} 
\vspace{-2ex}
\caption{The \transfer $\Twave_{\epsilon}$ for the density perturbations (log scale) as
a function of conformal time $\tilde{\eta}$. The family of solutions covers the
wave numbers range $\log(\tilde{k})\in [-1, 5]$ i.e. the lowest curve on the
diagram refers to the perturbation scale ten times larger than the sound horizon
distance, while the top one to a perturbation scale $10^{-5}$ times smaller.
Point $\tilde{\eta}=1$ represents the transition from the radiation to the
matter dominated universe.}
     \label{fig1a}  
     \end{figure}

\begin{figure}[h]
\centering 
\psfig{file=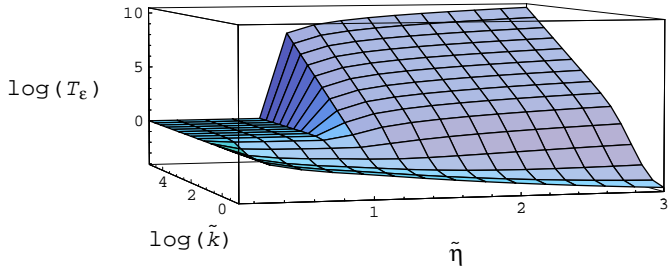,width=88mm 
}
\vspace{-2ex}
\caption{The density \transfer $\Twave_{\epsilon}$ 
as a function of both the conformal time $\tilde{\eta}$ and the
wave number $\tilde{k}$ ($\tilde{\eta}=1$ - the transition from the radiation to the
matter dominated universe).}
     \label{fig1b}  
     \end{figure}

To express the perturbation enhancement relative to its initial amplitude we introduce the ratio
	\begin{equation}
	\label{eq41}
{{\Twave}}_{\epsilon}(\tilde{\eta},\tilde{\eta}_i,\tilde{k}) =
\frac{p_k(\tilde{\eta})}{p_k(\tilde{\eta}_i)}
= \frac{\T_{\epsilon}(\tilde{\eta},\tilde{k})}{\T_{\epsilon}(\tilde{\eta}_i,\tilde{k})}.
	\label{stf}
	\end{equation}
The so called {\it transfer function\/} $\delta/\delta_i $ is often used in a
similar context. Fitting to this traditional name we will call $\Twave(\eta,k)$
the {\it \transfer}. Contrary to $\delta/\delta_i $ the {\it \transfer} 
$\Twave$ is real and invariant under unitary transformations of Fourier bases.
Plotted on a logarithmic scale (Fig.~\ref{fig1a}, Fig.~\ref{fig1b}) it shows that the
perturbations evolution is highly sensitive to $\tilde{k}$. In both Fig.~\ref{fig1a}, and
Fig.~\ref{fig1b} we start from $\eta_i=0.1~\eta_{\tsigma}$ ($\tilde{\eta}_i=0.1$) to
evaluate perturbation up to $\eta = 3~\eta_{\tsigma}$ ($\tilde{\eta}=3$).

As already mentioned, the large scale inhomogeneities (those, which are larger than the local sound
horizon $\tilde{k}\ll1$) decay in the radiation era ($\tilde{\eta}<1$). In
this case,   the  term $\frac{1}{\tilde{k}^2\tilde{\eta}^2}$ is dominant in the
\stf $\T_{(\epsilon,1)}(\tilde{\eta},\tilde{k})$ and  strongly
decreases with time. Decay of the large scale component
is a generic feature of acoustic noise on the expanding radiation dominated
homogeneous background. In other words, the homogeneity of the radiation filled
universe is a stable property at least as long as the generic perturbations
are taken into account. Similar phenomenon of large-scale wave
extinction is observed in scalar field theory (\cite{Stebbins&Veerarghavan}).
Although the large-scale inhomogeneities change substantially during the radiation
era, their response to the change in the equation of state is very weak. Their
wave character vanishes at $\eta_{\tsigma}$, but the spatial spectrum itself is
insensitive to the transition (see Fig.~\ref{fig1b}).

The low scale perturbations produce a different scenario. Their amplitudes kept
constant during the radiation dominated-epoch to increase by several orders of
magnitude at the transition. For inhomogeneities of galactic scale 
($M= 10^{11}M_s$) the \stf $\T$ is of $10^8$, which means a $10^4$ times
amplitude enhancement. The pressure discontinuity excites low-scale
perturbations and their contribution to the spatial-spectrum becomes dominant
shortly after the transition.

\begin{figure}[h]
\centering 
\psfig{file=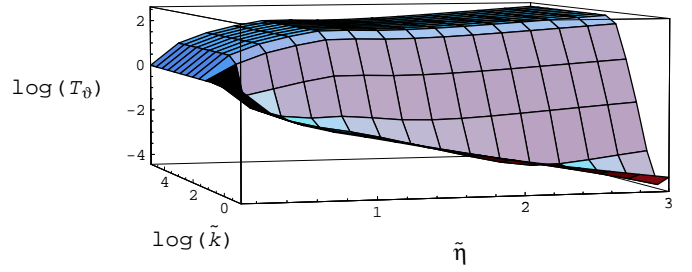,width=88mm 
} 
\vspace{-2ex}
\caption{ The expansion {\it spectrum transfer} $\Twave_{\vartheta }$
as a function of the conformal time $\tilde{\eta}$ and the
wave number $\tilde{k}$.
}
     \label{fig2a}  
     \end{figure}

\begin{figure}[h]
\centering 
\hspace*{10mm}\psfig{file=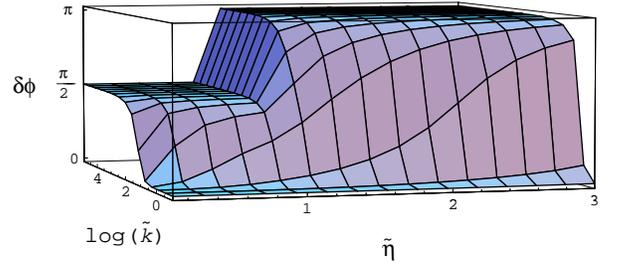,width=78mm 
}
\vspace{-2ex}
\caption{
Phases of perturbations.}
     \label{fig2b}  
     \end{figure}

As opposed to the density perturbations, the velocity magnitude does not change
significantly at the transition. An analogue of the transfer function
${{\Twave}}_\vartheta(\tilde{\eta},\tilde{k}) = \T_\vartheta(\tilde{\eta},
 \tilde{k})/\T_\vartheta(\tilde{\eta}_i,\tilde{k})$ constructed for the
expansion spectrum is shown in the Fig.~\ref{fig2a}. The expansion contrast increases
systematically during the radiation era and smoothly enters the matter era,
keeping nearly the same growth rate before and after the transition. A
substantial change in the velocity field relates to its phase. Figure Fig.~\ref{fig2b}
shows the difference in arguments of the density and expansion modes
$\delta\phi= \arg(\ue)-\arg (\uv)$. In the radiation era large frequency modes,
(both $\ue$ and $\uv$) are shifted to each other by $\pi/2$ (compare also (\ref
{eq39.1}) and (\ref{eq25})). This effect is the generic perturbation property in
the radiation dominated fluid (\cite{Ellis&Hellaby&Matravers}), and is an artefact
of their acoustic character. Similar relation of phases can be also obtained on
the ground of binding energy analysis (\cite{Liang&&}).  After the transition the
arguments of the density and velocity modes differ by $\pi$ so their maxima are
anti-correlated. Smallest expansion occurs in the regions of highest density.
Decreasing modes have decayed and fluid flow becomes potential
(\cite{Peebles&&}). (Slightly more complex is the behaviour of the low frequency
modes. Solutions like that hardly achieve the phase of potential flow (the
argument shift is different from $\pi$).)

>From the hydrodynamic point of view, the transition from the radiation
to the matter- dominated epoch breaks down the acoustic approximation.
The sound velocity instantly falls from $v=1/\sqrt{3}$ to zero, and in
consequence, the fluid velocity formally becomes greater than the sound
velocity at each point in space. (In this way the structure formation
has some aspects characteristic for formation of acoustic shocks -
compare \cite{acoustic1, acoustic2}.) The growth of inhomogeneities is
based on acoustic instability -- the self-gravitation processes do not
switch on until the linear regime fails. In both cases --- the density
and expansion fields --- the \transfer monotonically increases with the
wave number~ $k$. In particular the phenomenon of acoustic peaks is
absent\footnote{The same effect appears in more sophisticated
transition models investigated numerically (\cite{Press&Vishniac}), if
they include the complete basis of solutions. On the other hand, as
shown in (\cite{ Voglis&&}), peaks may appear in a pure radiation--filled 
universe model without evoking complicated recombination
processes, if one limits to growing modes (standing waves) with
specific phase correlation. These phenomena have also been discussed
in (\cite{ Fang&Wu, Riazuelo&Deruelle})}. The absence of peaks  is
characteristic feature of the random field composed of the
statistically independent moving waves (\cite{Grishchuk2&&}). {The
\transfer may substantially change in the small 
$k$-regime when the universes undergoes more than one phase transition 
(some classical ones, like
the change in the sound velocity, or semiclassical like the transition
from inflationary to radiation dominated universe). In these cases the
squeezed states in the acoustic field appear on large scales, which are
an alternative explanation of the CMBR temperature spectrum 
(\cite{Bose&Grishchuk}).}

\section{Growing modes versus complete solutions}

When considering instantaneous transition to the matter dominated universe 
authors limit
themselves to the large scale perturbation regime. In this regime {\em the growing mode amplitude after the phase
transition is entirely dominated by that before the phase transition and
there is no chance of generating a growing mode out of a decaying mode} 
(\cite{ Kodama&Sasaki}). Although the short perturbations 
do not produce the same scenario\footnote{{ \em
(...) the general relation among the amplitudes (...) would be too
complicated to extract any physical information out of it} 
(\cite{Kodama&Sasaki}).}  the decaying modes
of any length-scale are commonly neglected
at the beginning of the matter dominated era.

\begin{figure}[h]
\centering 
\hspace*{10mm}\psfig{file=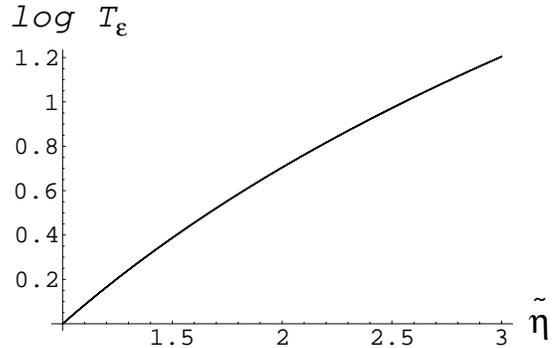,width=78mm 
}
\vspace{-2ex}
\caption{
The density \transfer for perturbations composed of pure growing modes.}
     \label{fig3}  
     \end{figure}

By abandoning all the decaying terms in the formula (\ref{eq37.2}) one obtains
$\T_{\epsilon}\propto (1+\tilde{\eta})^4$, and consequently, the spatial spectrum of
the density contrast
\begin{equation}
p_k(\tilde{\eta})=
{{\Twave}}_{\epsilon}(\tilde{\eta},\tilde{\eta}_i)p_k(\tilde{\eta}_i)
\label{grow1}\end{equation}
divides into time dependent 
\begin{equation}
{{\Twave}}_{\epsilon}(\tilde{\eta},\tilde{\eta}_i) =
\left(\frac{1+\tilde{\eta}}{1+\tilde{\eta_i}}\right)^4
\label{grow2}\end{equation}
and the scale dependent $p_k(\tilde{\eta}_i)$ (given at some $\tilde{\eta}_i>1$). 
The \transfer ${{\Twave}}_{\epsilon}$ drawn in the
logarithmic scale is presented in the Fig.~\ref{fig3}. 
${{\Twave}}_{\epsilon}$ does not depend on
$k$, therefore the perturbation is amplified in the scale-independent way. This
scale-independence is commonly attributed to the dust-filled universe (p=0), but
actually it requires more restrictive conditions: the perturbation must be
solely composed of growing modes.

While the pure growing modes increase ten times
in the interval $\tilde{\eta}\in(1,3)$, the mixture of growing and decaying
low scale modes is enhanced by several orders of magnitude 
(compare Fig.~\ref{fig1a} and Fig.~\ref{fig3}). 
The enhancement of this mixture depends on the
wave number $k$. Substantial magnification of the low scale modes 
(Fig.~\ref{fig1a}) means that the low-scale
structures may enter the nonlinear regime first. 

When the decaying modes are taken into account the scale-independence breaks
down. Decaying modes "remember"  perturbations'  past. In our case  the
admixture of decaying modes defined by the joining conditions (\ref{eq21},
\ref{eq22}) is an imprint of the acoustic (travelling wave) character of the
density perturbations in the radiational era. Consequently, the radiational era
affects the structure formation processes occurring latter on, when matter
dominates. Instability of perturbations is the genuine feature of the
entire cosmological evolution rather than the property of separate
cosmological epochs.

\section{Observables and observational constraints}

Although our primary goal is to discuss the role of decaying solution in the
formation of the small-scale structures, it is worth to check the model
consistence with data presently available for larger scales.

Data coming from the large galaxy surveys (SDSS, 2dFGRS and others) probe the
large-scale mass distribution at late epochs, while the WMAP experiment reports
the density fluctuations at the last scattering surface. These two categories of data potentially
allow reconstructing the \transfer i.e. the scale of inhomogeneity
enhancement after decoupling.

\begin{figure}[h]
\centering 
\psfig{file=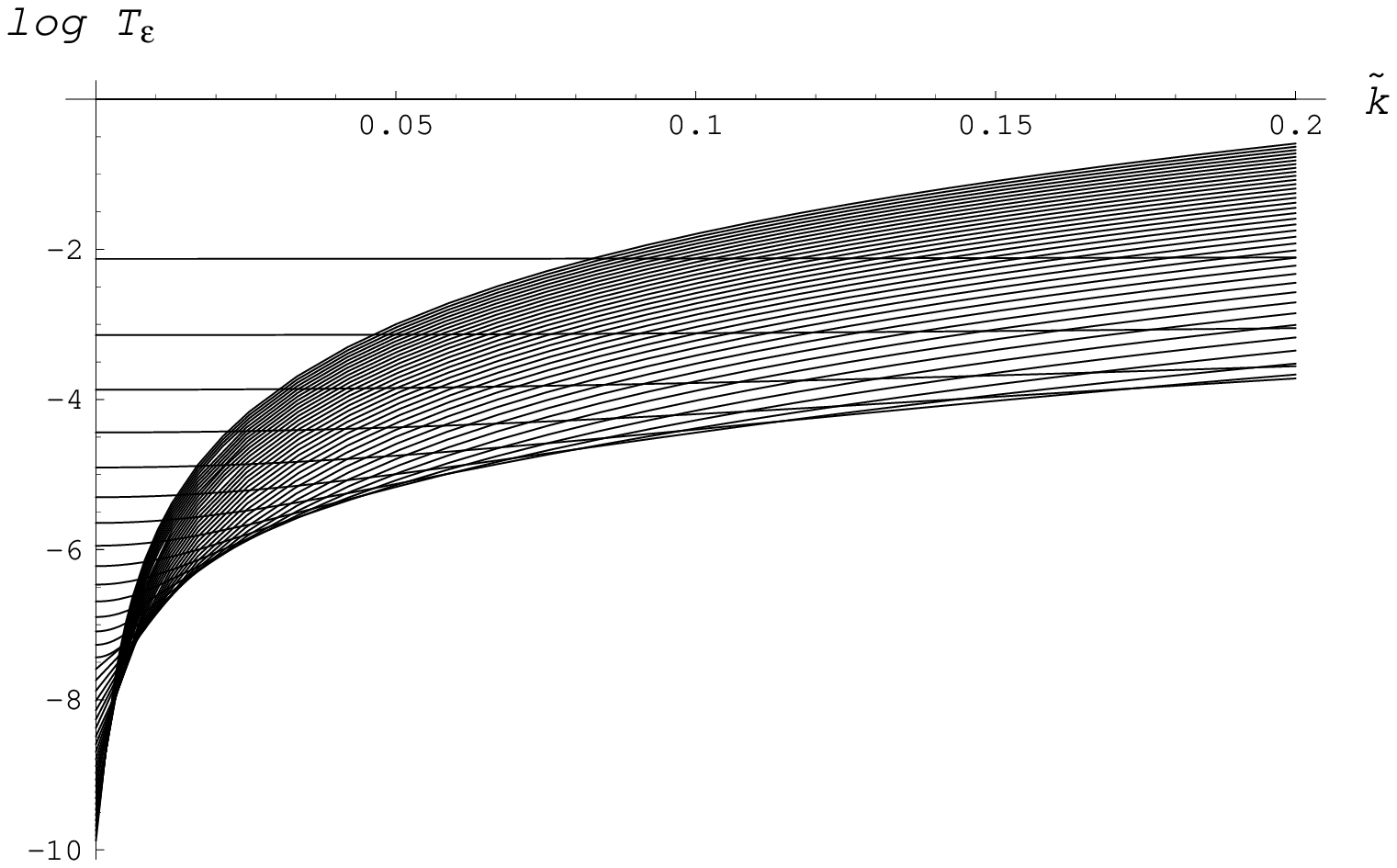,width=88mm 
}
\vspace{-2ex}
\caption{The \transfer ${\Twave}_{\epsilon}$ as a function of $\tilde{k}$ for 
diffrent $\eta\in(2,31)$}
     \label{Fig4}  
     \end{figure}

To examine the \transfer (\ref{stf}) we focus on the LSS spectral estimations,
with the scale range of some 100 Mpc, that at the same time overlaps the right
end of the range ($k \sim$ few ~$10^{-1}Mpc^{-1}$) probed by WMAP experiment. As
an example we take two close length-scales,  $k \sim 0.1~ Mpc^{-1}$ and $k \sim
0.05~ Mpc^{-1}$, for which the values of $\delta$  contrast (Fig. 3  in \cite{Wu&Lahav&Rees}) are
respectively, $\sim 10^{-2}$ and $\sim 10^{-3}$. These scales correspond to
$l\sim 1800, 700$ in the CMB spectrum and may be considered as the intermediate length--scales. 
The \transfer  ${{\Twave}}_{\epsilon}$ in this scales is given in the
Fig. \ref{Fig4}. The curves are drawn for the constant $\eta$--surfaces between $\eta \sim 2$ (the
lowest) and $\eta$ -- 31 --- corresponding roughly to redshift $\sim 0$. The
increase of the model \transfer (\ref{stf}) are roughly of the order of  $10^{3}
$ and $10^{2}$ at the scales under discussion, and reproduce the current density
contrast, from the initial density contrast of the order $\sim 10^{-5}$. In
spite of the high idealization, we obtain the rough agreement with observations
in $k \leq 1 Mpc^{-1} $ regime.

Substantially more difficult is to view the \transfer in context of extremely
low, or extremely large scales. In the large scale limit the \transfer
(\ref{stf}) becomes insensitive to changes in the equation of state. This
confirms that the large scale perturbations are ``fossils" of the remnant past,
what is interesting in context of their quantum or semiclassical
origin. Recently reported (e.g. \cite{Oliveira-Costa}, \cite{Hannestad&&})
discrepancy between the WMAP observational spectrum and that coming from the
inflationary paradigm for scales $k\leq 0.001~Mpc^{-1}$ challenges some revision
of the initial perturbations spectrum theory. On the other hand the
interpretation of measurement also deserves careful examination. The
perturbations and the resulting temperature fluctuations are commonly related to
each other by means of Sachs-Wolfe formula - an integral over gravitational
potential performed along the photon path (\cite{Sachs&Wolfe}, \cite{Peebles&&}).
The formula is true in absence or neglect of the decaying modes at the last
scattering surface (\cite{Sachs&Wolfe}). For the density perturbation matched to
the acoustic field the decaying modes cannot be neglected, particularly not the low
multipoles, where discrepancies become substantial\footnote{To neglect decaying modes
one needs $(b_2)^2<<(a_2)^2$.
After employing (\ref{eq28}) this condition reads 
$1+40/(3 k^2 {\eta_\Sigma}^2)+100/(k^4 {\eta_\Sigma}^4)<<1$, 
and is false for any $k$. Particularly, in the $k\rightarrow 0$ the error become
arbitrarily large.}. To understand properly the measured $C_l$ coefficients at
low-l limit we need the exact not approximate formula for the temperature
fluctuation. (In the discussed model the exact formula can be found, yet it is
out of the scope of this paper.)

Violent formation of fine scale structures in the transient epoch, what is the
characteristic feature of the model, enables inhomogeneities to enter nonlinear
regime soon after decoupling. The model qualitatively supports observations of
highly developed structures at high redshifts {\em z}. However, WMAP angular
resolution limit rules out any quantitative estimations for the structures less
than $10^4 ~M_{gal}$. Expecting more relevant data from the forthcoming Planck
mission we anticipate, on strength of the \transfer (\ref{stf}), that the CMBR
temperature fluctuations in the fine-scale regime may have relatively low
amplitude.

\section{Summary}

Generic density perturbations in the radiation dominated universe propagate in
the same way as sounds propagate in air or electromagnetic waves in vacuum. As
shown by Sachs and Wolfe (\cite{Sachs&Wolfe, White_PC&&}) perturbations form
waves travelling with the same speed $v=\frac{1}{\sqrt{3}}$ independently of
their scales or profiles, hence gravitationally bound structures cannot form.
Perturbations do not self-gravitate in the linear regime, so gravity may affect
their evolution merely by affecting the dynamics of the homogeneous background.
The wave character of the density perturbations (independent of their scales) is
confirmed in the Hamiltonian formalism (\cite{Lukash&&,Chibisov&Mukhanov}).

The hyperbolic type of propagation equation requires appropriate perturbation
statistics, where acoustic waves travelling in different directions are
statistically independent. This kind of statistics form a classical limit for
quantum theories (\cite{Lukash&&, Chibisov&Mukhanov, Grishchuk2&&}) and is
compatible with the gravitational waves theory 
(\cite{Abbott&Harari,Allen&&,Allen&Romano, Maggiore&&}). 
Probability and appropriate expectation values may
depend on the wave frequency, but not on the direction of propagation, neither
the wave phase at any time or position. A random choice of plane waves
guarantees that the perturbations and their canonical momenta are statistically
independent and uncorrelated quantities at any time. This finally results in the
stability of homogeneous expanding environment. Perturbations larger than the
sound horizon  decay during the radiation era, while those well inside the
horizon keep their magnitude constant in time. This property, although
contradicting Jeans conjecture, confirms results obtained in other gauges:
synchronous (\cite{Grishchuk1&&, Golda&Woszczyna1}), longitudinal
(\cite{Mukhanov&Feldman&Brandenberger} part III) and in Hamiltonian formalism
(\cite{Chibisov&Mukhanov}). Similar decrease in amplitude occurs for the large
scale component of the scalar field (Stebbins \& Veerarghavan 1993).

Cosmic structure formation naturally occurs at the transition from the
radiational to matter domination era. To match the acoustic field at the
decoupling, the growing and decaying modes contribute in the short wave limit
with nearly opposite phases, and therefore, compensate each other at~$\Sigma$.
After the transition both modes ``decouple" and the resulting superposition grows
explosively. Physically it means that the pressure decay from $\P=\epsilon/3$ to
$\P=0$  excite perturbations much lower than horizon scale, while leave
untouched the amplitude of those, which substantially exceed horizon\footnote{
This behaviour is opposite to that of tensor perturbations. A similar transition
in the equation of state will amplify large scale gravitational waves not
affecting the small scale ones.}. For growth of inhomogeneities the break down
of the acoustic approximation is responsible, i.e. the same group of physical
phenomena that may excite shock waves in the interstellar medium. This kind of
instabilities cannot be described by any formalism, which {\it a' priori}
neglects the role of decaying modes, no matter how realistic are the models for
physics of recombination and decoupling, which are used. 

This is obvious that the sharp transition between cosmological eras is not a
realistic model for decoupling or recombination. In reality these phenomena are
continuous, take some cosmologically substantial time, and involve a number of
complex physical processes. The recombination and decoupling do not coincide.
One can hardly expect that the realistic situation can be described by simple
analytical solutions as presented in this paper. Eventually one have to apply
numerical codes involving multi fluid hydrodynamics or magnetohydodynamics.
Still, the problem of the initial state remains. The physical meaning of the
obtained numerical results strongly depends on their stability against initial
condition and on the physical relevance of the initial state assumed at early
epochs. Most of hydrodynamical codes are ready to work with travelling waves,
therefore, the complete numerical analysis of random acoustic fields in the
expanding universe-- without neglecting {\it a' priori} the role of `decaying
modes' -- is basically possible. On the other hand, simple but nontrivial
analytic solutions presented in this paper may easily be used to verify numeric
procedures.

\section*{Acknowledgements}

We thank Zdzis{\l}aw Golda for valuable critical comments. 
This work was partially supported by State Committee for
Scientific Research, project  \grant.

\renewcommand{\theequation}{\Alph{section}\arabic{equation}}
\setcounter{section}{1}
\setcounter{equation}{0}

\def\references{\leftskip24pt\parindent-24pt\footnotesize}
\references



\begin{thebibliography}{}
\bibitem[Abbott \oraz Harari 1986]{Abbott&Harari}Abbott, F.L., \oraz Harari, D.D., 1986, Nucl.Phys., B264,487
\bibitem[Allen et al. 2000]{Allen&Flanagan&Papa}Allen, B., Flanagan, E., \oraz Papa, M.A., 2000,\\ Phys.Rev., D61, 24024
\bibitem[Allen 1996]{Allen&&}Allen, B., 1996, (gr-qc/9604033)
\bibitem[Allen \oraz Ottewill 1997]{Allen&Ottewill}Allen, B., \oraz Ottewill, A.C., 1997, Phys. Rev., D56, 545 
\bibitem[Allen \oraz Romano 1999]{Allen&Romano}Allen, B., \oraz Romano, D., 1999, Phys. Rev., D 59, 102001 
\bibitem[Anderson 1971]{Anderson&&} Anderson, T.W., 1971, {\em The statistical analysis of time series}, 
					(John Wiley \& Sons, New York)
\bibitem[Bardeen 1980]{Bardeen&&}Bardeen, J.M., 1980, Phys. Rev., D22, 1882
\bibitem[Birrell \oraz Davies 1982]{Birrell&Davies}Birrell, N.D., \oraz Davies, P.C.W., 1982, 
         {\it Quantum Fields in Curved Space,} (Cambridge University Press, Cambridge)
\bibitem[Brandenberger at al. 1983]{Brandenberger&Kahn&Press}Brandenberger, R., Kahn, R., \oraz Press, W.H., 1983, Phys. Rev., D28, 1809
\bibitem[Bose \oraz Grishchuk 2002]{Bose&Grishchuk} Bose, S., \oraz Grishchuk, L.P., 2002, Phys.Rev., D66, 43529
\bibitem[Chibisov \oraz Mukhanov 1982]{Chibisov&Mukhanov}Chibisov, G.V., \oraz Mukhanov, V.F., 1982, MNRAS, 200, 535
\bibitem[Darmois 1927]{Darmois&&}Darmois, G., 1927, {\em M\'{e}morial de Sciences
	Math\'{e}matiques, Fasc XXV, Les Equations de la Gravitation
	Einsteinienne}, (Gauthier-Villars, Paris) 
\bibitem[Deruelle \oraz Mukhanov 1995]{Deruelle&Mukhanov}Deruelle, N., \oraz Mukhanov, V.F., 1995, Phys. Rev., D52, 5549
\bibitem[Ellis \oraz Bruni 1989]{Ellis&Bruni}Ellis, G.F.R., \oraz Bruni, M., 1989, Phys. Rev., D40, 1804
\bibitem[Ellis et al. 1990]{Ellis&Bruni&Hwang}Ellis, G.F.R., Bruni, M., \oraz Hwang, J., 1990, Phys. Rev., D42, 1035
\bibitem[Ellis et al. 1990]{Ellis&Hellaby&Matravers}Ellis, G.F.R., Hellaby, C., \oraz Matravers, D.R., 1990, ApJ, 364, 400
\bibitem[Fang \oraz Wu 1996]{Fang&Wu}Fang, L.Z., \oraz Wu, X.P., 1996, astro-ph/9601087
\bibitem[Field \oraz Shepley 1968]{Field&Shepley}Field, G.B., \oraz Shepley, L.C., 1968, Ap{\&}SS, 1, 309
\bibitem[Frieman \oraz Turner 1984]{Frieman&Turner}Frieman, J.A., \oraz Turner, M., 1984, Phys.Rev., D30, 265
\bibitem[Golda \oraz Woszczyna  2001]{Golda&Woszczyna2}Golda, Z., \oraz Woszczyna, A., 2001, Class.Quantum Grav., 18, 543, (gr-qc/0002051)
\bibitem[Golda \oraz Woszczyna 2001]{Golda&Woszczyna1}Golda, Z., \oraz Woszczyna, A., 2001, J. Math. Phys., 42, 856
\bibitem[Grishchuk 1974]{Grishchuk0&&}Grishchuk, L.P., 1974, JETP,	67, 825
\bibitem[Grishchuk 1994]{Grishchuk1&&}Grishchuk, L.P., 1994, Phys. Rev., D50, 7151
\bibitem[Grishchuk 1995]{Grishchuk2&&}Grishchuk, L.P., 1995, (gr-qc/9511074)
\bibitem[Grishchuk 1996]{Grishchuk&&}Grishchuk, L.P.,  1996, (gr-qc/9603011)
\bibitem[Grishchuk 1998]{Grishchuk3&&}Grishchuk, L.P., 1998, (gr-qc/9801011)
\bibitem[Hannestad 2003]{Hannestad&&}Hannestad, S., 2000 (astro-ph/0311491)
\bibitem[Hawking \oraz Ellis 1973]{Hawking&Ellis}Hawking, S.W., \oraz Ellis, G.F.R., 1973, {\it The Large Scale Structure of Space-Time}\\
        (Cambridge University Press, London)
\bibitem[Hu 1998]{Hu&&}Hu, W., 1998, Phys.Rev., D59, 21301
\bibitem[Hwang \oraz Vishniac 1991]{Hwang&Vishniac}Hwang, J., \oraz Vishniac, E.T., 1991, ApJ, 382, 363
\bibitem[Israel 1966]{Israel} Israel, W., 1966, {\em  Nuovo Cimento}, 44B, 1 
	and corrections in {\em ibid}, 48B, 463 
\bibitem[Jackson 1993]{Jackson&&}Jackson, J.C., 1993, MNRAS, 264, 729
\bibitem[Kodama \oraz Sasaki 1984]{Kodama&Sasaki}Kodama, H., Sasaki, M., 1984, Prog. Theoret. Phys. Suppl., 78, 1
\bibitem[Khoperskov \oraz Khrapov 1999]{acoustic2}Khoperskov, A.V., \oraz Khrapov, S.S., 1999, A{\&}A, 345, 307
\bibitem[Liang 1977]{Liang&&}Liang, E.T.P., 1977, MNRAS, 180, 117
\bibitem[Lichnerowicz 1955]{Lichnerowicz&&}Lichnerowicz, A., 1955, {\it Theories relativistes da la gravitation et de l'electromagnetisme,} (Masson, Paris) 
\bibitem[Loeve 1963]{Loeve&&}Loeve, M., 1963, {\it Probability theory}, (Princeton University Press, Princeton)
\bibitem[Lukash 1980]{Lukash&&}Lukash, V.N., 1980, Sov. Phys. JETP, 79, 1601
\bibitem[Lukash 1999]{Lukash99&&}Lukash, V.N., 1999, (astro-ph/9910009)
\bibitem[Lyth \oraz Mukherjee 1988]{Lyth&Mukherjee}Lyth, D.H., \oraz Mukherjee, M., 1988, Phys. Rev., D38, 485
\bibitem[Lyth \oraz Stewart 1990]{Lyth&Stewart}Lyth, D.H., \oraz Stewart, E.D., 1990, ApJ, 361, 343
\bibitem[Maggiore 2000]{Maggiore&&}Maggiore, M., 2000, ICTP Lecture, June 2000, (gr-qc/0008027)
\bibitem[Montenegro et al. 1999]{acoustic1}Montenegro, L.E., Yuan, C., \oraz Elmegreen, B.G., 1999, ApJ, 520, 592	
\bibitem[Mukhanov et al. 1992]{Mukhanov&Feldman&Brandenberger}Mukhanov, V., Feldman, H., \oraz Brandenberger,R., 1992, Phys. Rep.   215, 203
\bibitem[Oliveira-Costa et al. 2003]{Oliveira-Costa}
         Oliveira-Costa, A., Tegmark, M., Zaldarriaga, M., et al., 2003, (astro-ph/0307282)
\bibitem[Olson 1976]{Olson&&}Olson, D.W., 1976, Phys. Rev., D14, 327
\bibitem[Padmanabhan 1993]{Padmanabhan&&}Padmanabhan, T., 1993, {\it Structure Formation in the Universe}	\\
        (Cambridge University Press, Cambridge)
\bibitem[Parker 1972]{Parker&&}Parker, L., 1972, Phys.Rev., D5, 2905,
\bibitem[Peebles 1980]{Peebles&&}Peebles, P.J.E., 1980, {\it The Large Scale Structure of the Universe}, \\
                   Princeton Series in Physics, (Princeton, New Jersey)
\bibitem[Press \oraz Vishniac 1980]{Press&Vishniac}Press, W.H., \oraz Vishniac, E.T., 1980, ApJ, 236, 323
\bibitem[Riazuelo \oraz Deruelle 2000]{Riazuelo&Deruelle}Riazuelo, A., \oraz Deruelle, N., 2000, Annalen Phys., 9, 288
\bibitem[Sachs \oraz Wolfe 1967]{Sachs&Wolfe}Sachs, R.K., \oraz Wolfe, A.M., 1967, ApJ, 147,73
\bibitem[Sakai 1969]{Sakai&&}Sakai, K., 1969, Prog. Theoret. Phys., bf 41, 1461
\bibitem[Sobczyk 1991]{Sobczyk&&}Sobczyk, K., 1991, {\it  Stochastic differential equations,} (Kluwer Academic Publishers B. V.)
\bibitem[Stebbins \oraz Veerarghavan 1993]{Stebbins&Veerarghavan}Stebbins, A., \oraz Veerarghavan, S., 1993, Phys. Rev.  D48, 2421
\bibitem[Stone 2000]{Stone&&}Stone, M., 2000, Phys. Rev., D62, 1341
\bibitem[Stoeger at al. 1991]{Stoeger&Ellis&Schmidt}Stoeger, W.R., Ellis, G.F.R., \oraz Schmidt, B.G., 1991, Gen. Rel. Grav. 23, 1169
\bibitem[White 1973]{White_PC&&}White, P.C., 1973, J. Math. Phys., 14, 831
\bibitem[White 1992]{White&&}White, M., 1992, Phys. Rev., D46, 4198
\bibitem[Whitham 1974]{Whitham&&}Whitham, G.B., 1974, Linear end Nonlinear Waves, (Wiley,New York)
\bibitem[Woszczyna \oraz Kulak 1989]{Woszczyna&Kulak}Woszczyna, A., \oraz Kulak, A., 1989, Class. Quantum Grav., 6, 1665
\bibitem[Wu et al. 1999]{Wu&Lahav&Rees} Wu, K., Lahav, O., \oraz Rees, M., 1989, Nature, 397, 225
\bibitem[Voglis 1986]{Voglis&&}Voglis, N., 1986, A{\&}A, 165, 10
\bibitem[Yaglom 1961]{Yaglom&&}Yaglom, A.M., 1961, {\it Second order homogeneous random fields,}\\
        in Fourth Berkeley Symposium Volume II, ed. J.    (University of California Press 1961)
\end{thebibliography}
\end{document}